Comment on "New probing techniques of radiative shocks"
by C.Stehlé et al, Opt.Comm. **285**, 64 (2012)

Michel Busquet
*ARTEP,inc – Ellicott City, MD 21042, USA*

**Abstract.**

In this comment, we discuss the possibility of imaging the radiative precursor of a strong shock with a 21.2 nm soft x-ray laser probe and we analyze the data presented in C.Stehlé et al "New probing techniques of radiative shocks", (Optics Comm. 285, 64, 2012) in order to derive some estimation of the achieved resolution. We show that the presented results are inconclusive for the existence of a radiative precursor. Furthermore, our best estimation of cold and warm Xenon VUV opacities tells that 21.2 nm backlighting would not be able to probe this radiative precursor.

**1- Introduction**

In C.Stehlé et al, 2011 [1] (named hereafter "paper I"), the authors present results obtained on the PALS facility for a Radiative Shock launched by a laser ablated piston in the line of previous similar experiments [2-5]. This time the authors used the unconverted laser ($\lambda$=1.315 µm) but at the same laser irradiance of $10^{14}$ W/cm$^2$, not accounting for the suggested 25% off pointing error [6]. Some of these results were also presented, with images of better quality, in the proceeding of a PALS meeting [7] (name hereafter "paper II"). The shock is launched in a Xe-filled miniature shock tube (0.4 mm x 6 mm) by a laser-ablated piston. The piston is a gold-coated plastic foil, where the plastic facing the laser converts the laser energy into piston velocity, using the so-called rocket effect. The gold layer acts as an inertia reservoir (the foil areal mass does not change during the observed times) and aims also to block hard x-rays created by the laser interaction process. However it may not stop suprathermal electrons created by resonant absorption (or x-rays created by them) that was suppressed in previous experiments by using the third harmonic ($\lambda$=0.438 µm). Hot spots and significant preheating of the gas may result. If the shock velocity is large enough and the preheat of the gas not too strong, a radiative precursor is launched by the shock.

The goal of the described experiment is to measure at the same time the position of the hydrodynamic shock (i.e. the density jump) and of the radiative precursor (a temperature wave, with no bulk density increase). Two diagnostics are presented. The first one is a spatially resolved VUV absorption. The absorption spatial profile is obtained with an x-ray laser probe at $\lambda$=21.2 nm, thanks to a special target design derived from the one used in the previous experiment [8]. Since no measure of the achieved spatial resolution has been presented, the reader has to derive spatial resolution (along the tube axis, the pertinent one) from the published images [1,7]. Our estimation of the spatial resolution of the transmission curve is 0.4± 0.1 mm as the readers can infer from the following section. Paper I presents also time resolved measurement of the self-emission in the VUV and XUV range. From the geometry described in the paper, one can infer a 7 ns time resolution for one of the diode, and 17 ns for the other one (see Sec.3). Recorded profiles rather suggest a 15-20 ns resolution.

In the following sections, we discuss the possibility to image the precursor with a 21.2 nm XRL probe and we analyze the presented data to derive some estimation of the achieved resolution.

**2- Analyzing VUV absorption profile**

This absorption profile is obtained from a coherent VUV imaging using an astigmatic off-axis spherical mirror. The angle of incidence is unknown, it is probably between 0.2 and 0.5 radian. From the description given in paper I, the magnification is 8.2 and the focal length is 30 cm, therefore the image plane is located at (1+8.2) x 30 cm =276 cm from the mirror. Shielding of the self-emission of the target is obtained with a 0.5 mm diameter diaphragm located in the focal plane, therefore at 246 cm from the image plane, resulting in a very low aperture of f/4920 (Fig.1). As the pinhole size is smaller than the object to be imaged, the different parts of the object (along the tube axis) are imaged by different parts of the mirror. Slope and position errors from the ideal curve of the reflecting surface result in a downgraded spatial resolution, in addition to the speckle pattern probably due to reflectivity variation of the multilayer coating, clearly seen in Fig.6 of paper I and page 25 of paper II. Finally, because the mirror is off axis, tangential and sagittal focal lengths differ by 12 to 80 mm for angle of incidence of 0.2 to 0.5 radian, resulting in an out of focus blurring *in one direction*, of resp. 100 to 500 microns. Refraction by the strong gradient inherent to the shock should not downgrade the spatial resolution. Shock also might be non planar (at small scale) if the heating beam is not smoothed (the presence of a smoothing zone plate is not mentioned in paper I) and this would add to the blurring of the shock front. The penumbra zone (around x=1.8 mm in Fig.7 of paper I, see also Fig. 7 of the present comment) has a 0.4± 0.1 mm extension [9] and is claimed by the authors to be a trace of the precursor, but can result from all the mentioned source of blurring along the shock propagation direction. Note that the decrease of transmission on the right (from 1 to 0.5 mm in Fig.7 of paper I) may also be attributed to limited spatial resolution. With a good longitudinal resolution of the transmission curve, the shock would appear sharply, with a ten times stiff increase in density ($N_i=N_e/<Z>$ value derived from Fig.5 of paper I).

Furthermore, we compute the Xenon opacity with the state-of-the-art atomic physics code STA [10] and we found that Xenon haves an almost constant opacity (Fig.2) for temperatures from 3 to 30 eV and a density of 1.5 mg/cm3 (i.e. pressure of 0.3 bar). By comparing different highly ranked opacity codes [11] we estimate a theoretical "uncertainty" of ± 30%, and much less for the relative variation with temperature. The whole precursor, if it exists, would then have a constant transmission of 0.6±0.1, which is very close (Fig.3) to the cold Xenon transmission of 0.7 [12]. Therefore the 21.2 nm XRL probe would not be able to discriminate the precursor from the cold Xenon and would not give the observed 50% contrast (Fig.7 of paper I) between the cold Xenon and the claimed-to-be-warm Xenon.

In summary, the large penumbra zone is more probably due to the insufficient quality of the imaging system, the deconvolution process, and/or the non-uniformity of piston velocity. High resolution would reveal the stiff density jump in the shock. We present in Fig.4 the computed transmission obtained with a 0.5 mm spatial resolution and a model two steps absorption profile corresponding to _first step_ the shocked gas and dense piston (less than 100 microns apart) and _second step_ to the ablated piston. The convolution shows a profile similar to the measured one, which tends to confirm low spatial resolution of the transmission curve.

### 3- Time resolved self-emission

The theoretical time resolution of the recording diodes can be inferred from the geometry described in paper I. The diodes, located at 55 mm of the shock tube, have an aperture of 1 mm along the shock propagation axis and image a portion of the shock tube with a 0.175 mm (0.3mm for the "OBL" diode) slit located at 9.5 mm (resp. 17 mm) of the shock tube thus at 45.5 mm (resp. 38 mm) of the diode. This give a theoretical field of view (FOV) of extent $FOV_{PERP} = ( 1 \times 9.5/45.5 + 0.175 \times 55/45.5 )= 0.42$ mm for the "PERP" diode, which corresponds to a 7 ns time resolution using a shock velocity of 60 km/s. The OBL diode view the shock section with an angle of 20 degrees which add 0.4 mm x sin(20 deg) to the FOV which is then equal to $FOV_{OBL} = (0.4 \sin(20)+ 1 \times 17/38 + 0.3 \times 55/38 )= 1.02$ mm , and gives a 17 ns time resolution. From the presented time traces, one would find a time resolution of the whole system of 15-20 ns, as deduced from the rise times and decrease times since a measurement of the achieved resolution is missing. Using this resolution, the traces are consistent with the expected rise of emissivity when the laser ablation front crosses the diodes FOV, possibly blurred with the self-emission of the hydrodynamic shock, and followed by the decreased emission from the corona (Fig.5). Obviously, we present here a crude description of the time history of emission, but details would be washed out by the low resolution. We point out that the emissivity of the ablation front and of the corona is apparently omitted in the estimation presented in paper I. Note in passing that the strong peak at t=0 might be due to stray laser light, *or* to electromagnetic impulsive loading of the electric circuit through antenna effect.

## 4- Conclusion

Paper I examine the possibility of a novel idea of radiography using a coherent x-ray laser beam to analyze a strong shock propagating in a Xe-filled mini shock tube, with a possible radiative precursor. However, the presented results reveal merely a time or space resolved measure of the shadow cast by the piston (and the density jump of the shock less than 100 µm apart). Estimated low spatial resolution and other sources of longitudinal blurring prevent any further conclusions. The measurement of self-emissivity is not conclusive either. No evidence of any precursor can be inferred from these measures as it is claimed. A precursor may exist, but it would require an improved diagnostic system to be revealed. Moreover, as our present estimation of the opacity at 21.2 nm of warm Xenon is almost equal to the cold Xenon opacity, we believe that the precursor would not be probed with this technique.

Up to now, non-coherent hard X-ray radiography [13-17] has provided more information on the spatial structure of strong radiative shocks, with the same launching system and similar conditions (0.5 to 6 mg/cm3 and ~ 600 µm transverse dimension).

**Figure and figure captions**

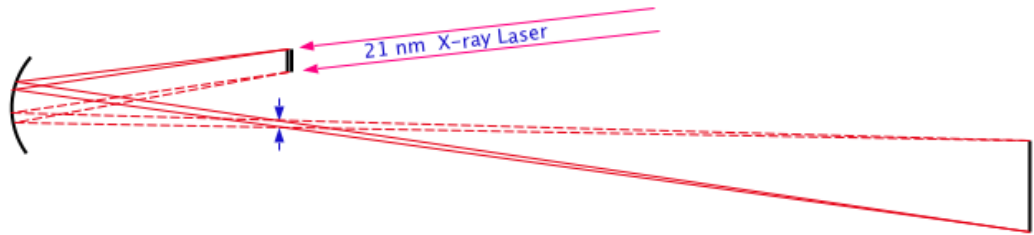

Fig.1 *Imaging system layout, here drawn for a magnification of 4, showing the shock tube to be imaged (drawn as a short thick line), the spherical off-axis imaging mirror, the apodizing diaphragm (blue arrows in this figure) and the image. Distances from object to mirror, from mirror to the 0.5 mm diaphragm (set at the focal spot of parallel rays), and from diaphragm to image plane, are deduced from the focal length and the magnifying ratio given in paper I (30 cm and 8.2). They are respectively 33.7 cm, 30 cm and 246 cm. The aperture of the optical system is then 0.05/246. Different positions of the mirror image different positionsd of the shock tube (imaging of each end is drawn with resp. plain line and dashes).*

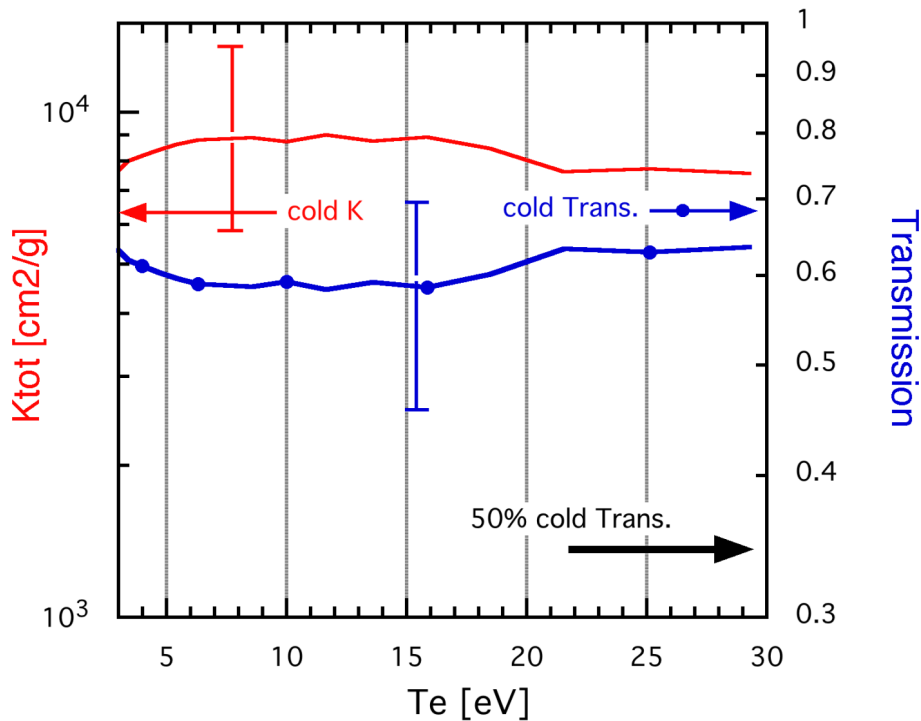

Fig.2 - *Computed opacity at $\lambda=21.2$ nm of 0.3 bar Xenon and transmission of 0.4 mm slab vs. temperature. Low contrast with cold opacity and transmission (arrows) is found. Black thick arrow is drawn for 50% of the cold Xenon transmission.*

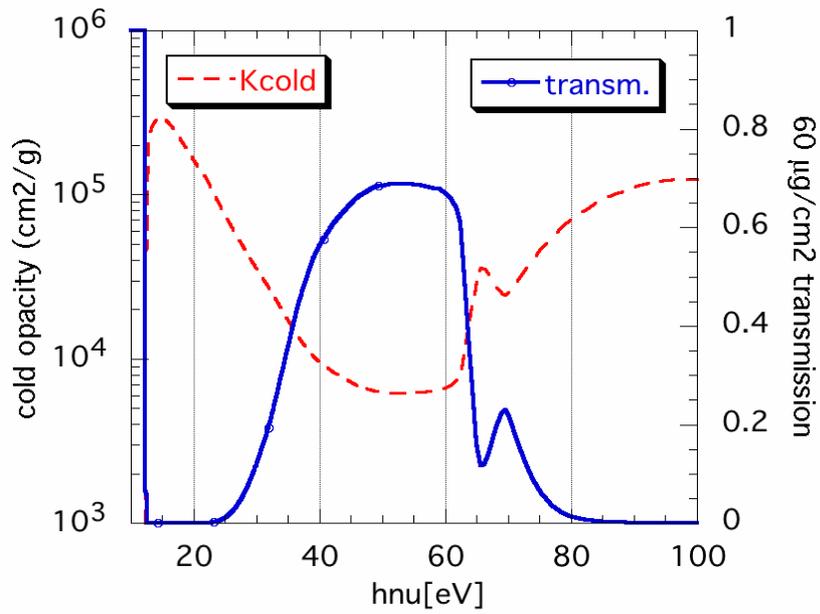

*Fig.3 - 0.3 bar Xe cold opacity [11] vs. photon energy and transmission of 0.4 mm Xe gas at 0.3 bar.*

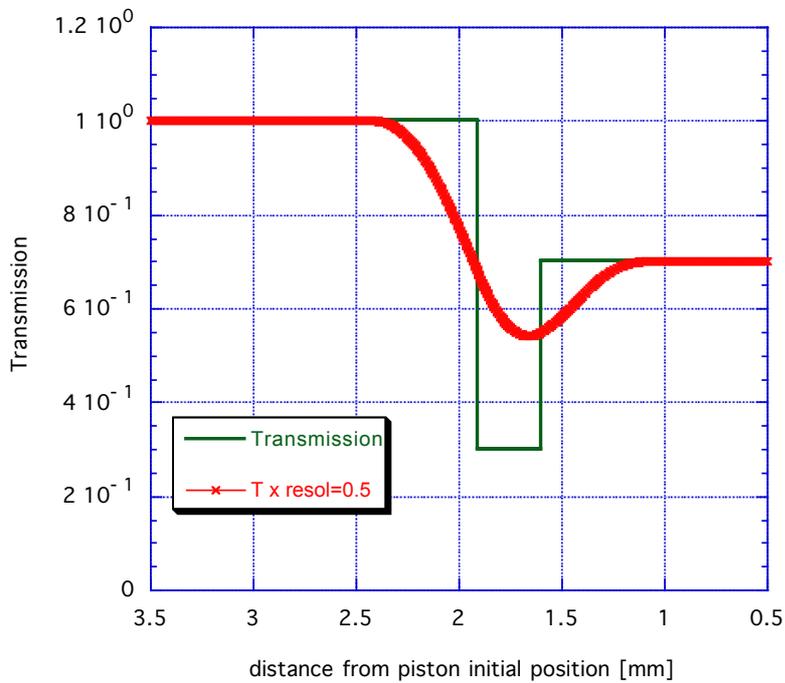

Fig.4 - *Spatial transmission given by a simple shock (green, thin line) convolved with a 0.5 mm spatial resolution (red, thick line).*

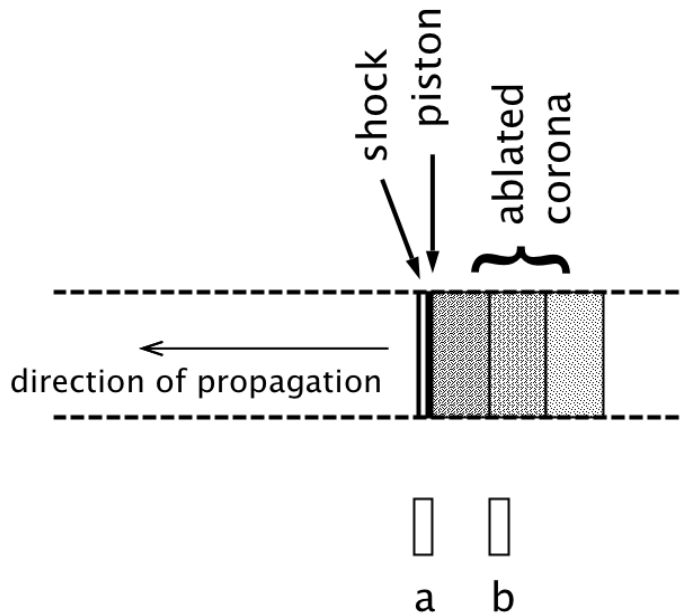

Fig.5 - *Schematic distribution of the emissivity in the piston velocity frame. The relative position of the diodes is given at approx. 20 ns (a) and 40 ns (b).*

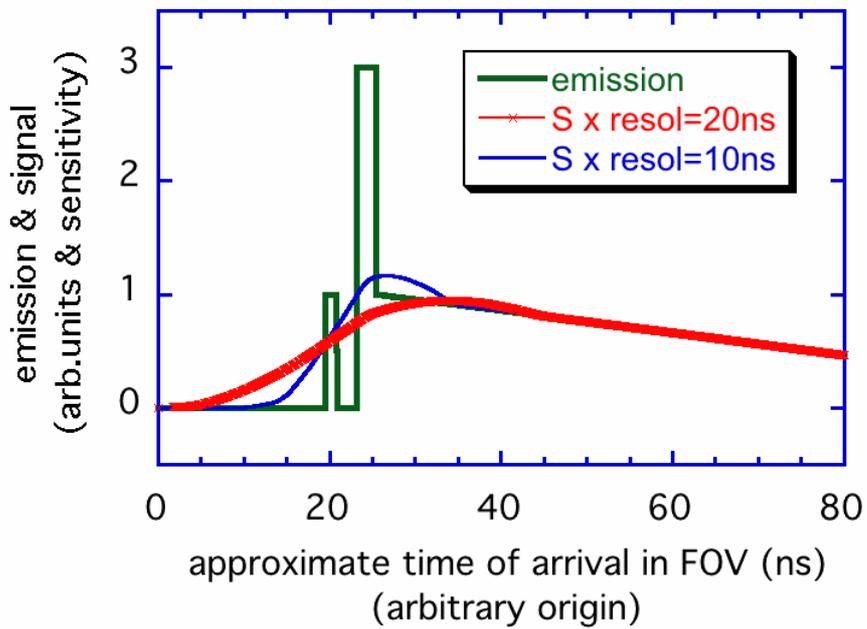

Fig.6 - *(Very) schematic time history of emission in FOV and signal with a 10 ns and 20 ns resolution.*

Fig.7 - *Transmission computed from half-tone images found in paper II gives a "shadow edge" extension of 0.3 to 0.4 mm. The transmission can be computed either from horizontal profile of "pixel to pixel image ratio" (blue thin line) using the ImageJ free software as speckle pattern appears reproducible, or from ratio of horizontal profiles of each images (red thick line, here after a bandpass FFT filtering).*

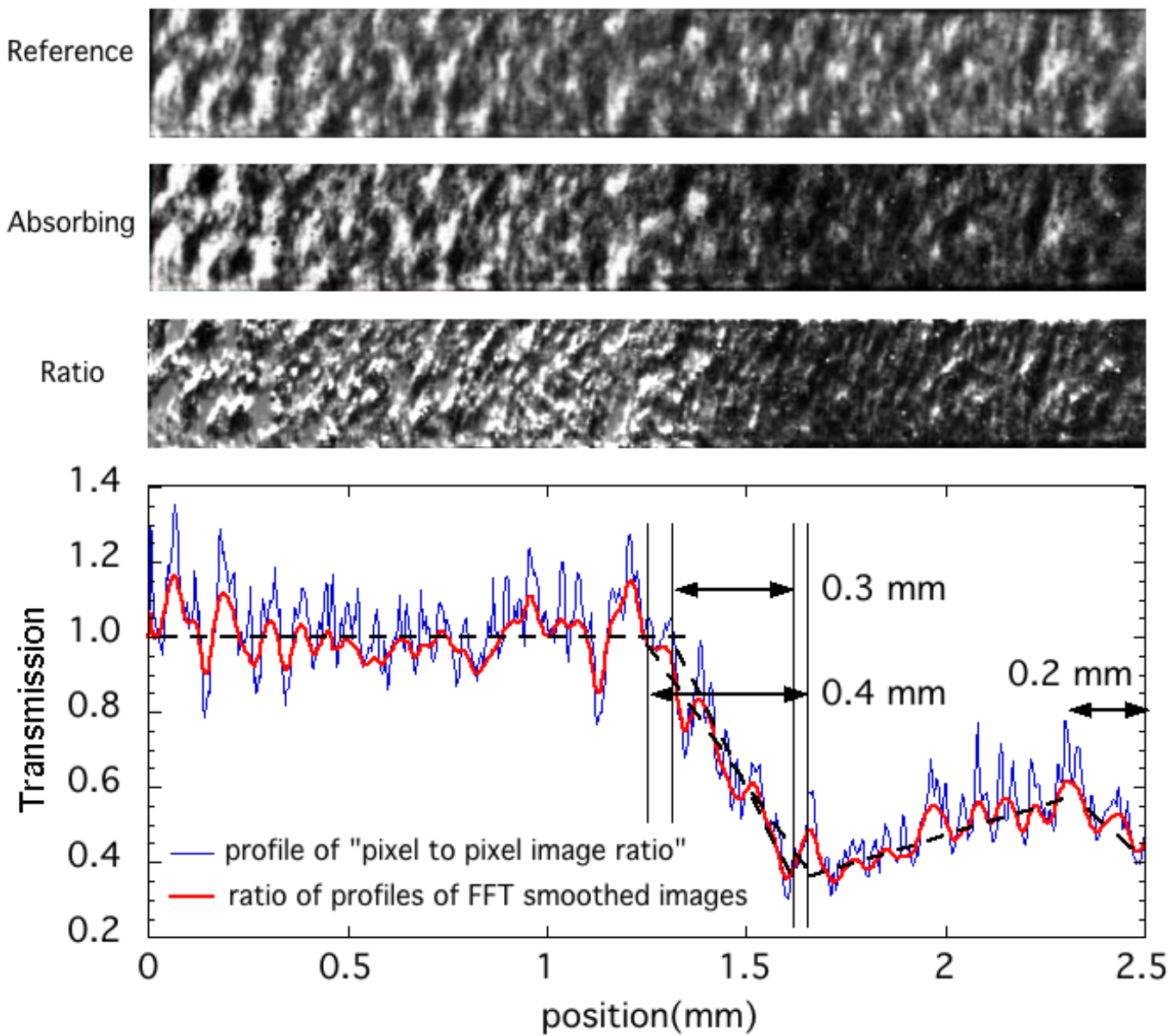